\documentclass[aps,pra,onecolumn,english]{revtex4}
\usepackage[T1]{fontenc}
\usepackage[latin9]{inputenc}
\usepackage{babel}
\usepackage{graphicx,amsmath,amssymb,color}
\usepackage[normalem]{ulem}
\usepackage{amsfonts}
\usepackage[toc,page]{appendix}
\usepackage[colorlinks=true]{hyperref}
\usepackage{latexsym}
\usepackage{amsfonts}
\usepackage{algpseudocode}
\usepackage{amsthm}
\usepackage{mathrsfs}
\usepackage{color,verbatim}
\usepackage{psfrag}
\usepackage[svgnames]{xcolor}
\usepackage{tikz}

\newcommand{\ket}[1]{|#1 \rangle}

\begin{document}

\title{An experimental test of search range in quantum annealing}
\author{Nicholas Chancellor and Viv Kendon}
\address{Department of Physics; Joint Quantum Centre (JQC) Durham-Newcastle \\ Durham University, South Road, Durham, DH1 3LE, UK}

\date{\today}

\begin{abstract}

We construct an Ising Hamiltonian with an engineered energy landscape such that it has a local energy minimum which is near to the true global minimum solution, and further away from a false minimum. 
Using a technique established in previous experiments, we design our experiment such that (at least on timescales relevant to our study) the false minimum is reached preferentially in forward annealing due to high levels of quantum fluctuations. 
This allows us to demonstrate the key principle of reverse annealing, that the solution space can be searched locally, preferentially finding nearby solutions, even in the presence of a false minimum. The techniques used here are, to the best of our knowledge, distinct from previously used experimental techniques, and allow us to probe the fundamental search range of the device in a way which has not been previously explored. 
We perform these experiments on two flux qubit quantum annealers, one with higher noise levels than the other. 
We find evidence that the lower noise device is more likely to find the more distant energy minimum (the false minimum in this case), suggesting that reducing noise fundamentally increases the range over which flux qubit quantum annealers are able to search. 
Our work explains why reducing the noise leads to improved performance on these quantum annealers. 
This supports the idea that these devices may be able to search over broad regions of the solution space quickly, one of the core reasons why quantum annealers are viewed as a potential avenue for a quantum computational advantage.

\end{abstract}

\maketitle

\section{Introduction and background}

Quantum annealing is an approach to quantum computing where optimization and other problems are mapped such that their solutions correspond to the ground and low energy states of a Hamiltonian, and quantum fluctuations are used to solve the problem.
%, has been a subject of much interest recently. 
Small proof-of-concept experiments have been performed based on a variety of real problems, with applications as diverse as cryptography \cite{Joseph20a}, design of radar waveforms \cite{coxson14a}, protein folding \cite{perdomo-ortiz12a}, air traffic control \cite{Stollenwerk19a}, scheduling \cite{crispin13a,Venturelli15a,Tran16a}, and hydrology \cite{omalley18a}. To the best of our knowledge, no scaling advantage has been observed on these devices for optimization, although recent work suggests one may be present for quantum simulation \cite{King2019b}.

Part of the motivation for interest in quantum annealing is the development of large scale superconducting flux qubit quantum annealers. 
While it is possible to build quantum annealers using other technologies such as trapped atoms \cite{Bernien2017a}, to date the largest and most technologically mature quantum annealing devices are the superconducting flux qubit quantum processing units (QPUs) produced by D-Wave Systems Inc., which are the subject of the experiments described here. 
In these devices, the low energy subspace of superconducting flux qubits is used to implement an Ising Hamiltonian described by a coupling matrix $J$ and a field vector $h$, which is subject to quantum fluctuations which flip single bits. 
The total effective Hamiltonian can be expressed as a transverse Ising Hamiltonian,
\begin{equation}
H(s)=-A(s)\sum_i \sigma^x_i+B(s)\left( \sum_{i,j \in \chi}J_{i,j}\sigma^z_i \sigma^z_j \sum_i+ h_i\sigma^z_i \right), \label{eq:H_tf}
\end{equation}
where $\sigma^{x(z)}_i$ is the Pauli $x$($z$) spin operator acting on qubit $i$, and $A(s)$ and $B(s)$ are time dependent real parameters. The annealing parameter $0 \le s\le 1$ controls the schedule such that in a traditional annealing protocol $s \propto t$. This parameterization is chosen so that reverse annealing protocols can be more easily described mathematically.
In this paper, we choose a convention where $J_{ij}$ and $h_i$ from Eq.~(\ref{eq:H_tf}) are dimensionless, while the control parameters, $A$, $B$ have dimensions of frequency. 
Coupling between pairs of qubits is restricted to a graph $\chi$ determined by the hardware. 
A quantum annealing protocol 
%consists of time dependent operation
is specified by the variation in time of the annealing parameters $A$ and $B$. 
%We conventionally define the parts constructed from $\sigma^z$ terms to be the ``problem'' Hamiltonian, 
The specific choices for the elements of $J$ and $h$ define the problem to be solved by finding the ground state, i.e., 
$H_{\mathrm{prob}}=\left( \sum_{i,j \in \chi}J_{i,j}\sigma^z_i \sigma^z_j \sum_i+ h_i\sigma^z_i \right)$.

Traditional quantum annealing is performed by starting the system in an equal positive superposition of all computational basis states with $\frac{A}{B}\gg1$ and varying the parameters continuously until $\frac{A}{B}\ll1$. 
In principle, in the absence of noise, and if the parameters are changed slowly enough, this protocol finds the solution state with a probability approaching one. 
In practice, the noise on the QPUs cannot be neglected, so the picture of how they operate is significantly more complicated. 
However, given that the dominant noise source on these devices is coupling to the low temperature substrate of the qubits, the net effect of noise is not always detrimental. 
This is emphasized by the result in \cite{dickson13a}, where it was found that thermal fluctuations can lead to a large enhancement of success probability over what is expected from a purely adiabatic protocol run over short timescales.  Evidence for quantum effects playing a role alongside thermal fluctuations is provided by \citeauthor{johnson11a} \cite{johnson11a} who demonstrate that fluctuations do not ``freeze out'' at low temperatures as one would expect for thermal fluctuations. 
\citeauthor{lanting14a} \cite{lanting14a} found experimental evidence which points to entanglement effects, and \citeauthor{boixo16a} \cite{boixo16a} showed that tunnelling rates on these QPUs agree well with a open quantum system model (which did not have any fitted parameters). The \citeauthor{boixo16a} study also disproved the idea that D-Wave devices are well described by (classical) spin vector Monte Carlo models, as suggested by \citeauthor{Shin2014a} \cite{Shin2014a}. 
More recent QPU devices have been used for quantum simulations of both transverse field spin glass systems \cite{Harris2018a} and Kosterlitz-Thouless phase transitions \cite{King2018a,King2019b}, showing good agreement with theoretical predictions. 

Given the good agreement between experiments and corresponding quantum open system models, within our work in this paper we implicitly assume that, while the system is noisy, quantum tunnelling effects still play an active role in the searches which the devices perform. Our work does not itself provide strong evidence of quantum effects, although it is consistent with this picture. 
We also do not claim that the quantum open system model provides a quantum speed up for solving optimization problems; our goal in the work is to better understand the mechanisms involved when annealing protocols are run on QPUs. 
Our work is a distinct advance on previous work to understand the effect of noise on D-Wave QPUs. Studying devices operating at different temperatures, or varying the temperature on the same device, is a commonly used tool to understand these devices. Important results obtained by varying temperature include \citeauthor{johnson11a} \cite{johnson11a}, who used temperature variation to show that temperature dependence of tunneling rates disappeared at low temperatures as would be expected from quantum processes, but not thermal ones. \citeauthor{Marshall17a} \cite{Marshall17a} examined the effect of temperature on thermal sampling, following on from \cite{chancellor16b}, which achieved the same effect by globally scaling the problem Hamiltonian. Furthermore, \cite{dickson13a} varied the temperature to explore thermal effects in tunneling and found a non-monotonic dependence of success probabilities, consistent with their model of thermally enhanced computation. Meanwhile, \cite{King2018a} and \cite{Harris2018a} used temperature controls to experimentally probe the phase diagrams of spin systems. 

In contrast, instead of varying temperature, we study the effect of a re-engineered chip to reduce the strength of the bath coupling (leading to reduced coupling to mid and broad band noise as characterized in \cite{low_noise_whitepaper}).  The only other work we are aware of which presents a performance comparison between more and less noisy devices is a white paper published by D-Wave systems around the time of the release of a lower noise QPU \cite{noise_reduction_spin_glass_whitepaper}.  While this white paper shows that performance is, indeed, improved by reducing the noise, it does not investigate the cause of the improved performance. The experiments we present in this paper are the only work we are aware of which experimentally probes the reason for improved performance in QPUs when noise coupling is reduced. This is a crucial question for understanding the role noise is playing. The main impact of bath temperature in an open quantum system model is determining the final equilibrium distribution of the system. Reducing noise coupling has a fundamentally different effect, by controlling how long the system remains coherent, and how fast the bath can add or remove energy.
Since, unlike temperature, bath coupling strength is not as readily controllable without manufacturing multiple devices, its effects on the dynamics has not been systematically studied on D-Wave QPUs.  Our work is to our knowledge the first step in that direction.

\subsection{Reverse annealing protocols}

Since late 2017, a new protocol has become available to users of D-Wave devices, known as reverse annealing \cite{reverse_anneling_whitepaper}. 
The QPU is initialized in a user defined computational basis state, with $\frac{A}{B}\ll1$.
The parameters are then changed to $\frac{A}{B}\sim 1$ to allow quantum fluctuations. 
Finally, the parameters are switched back to $\frac{A}{B}\ll1$, so each qubit can be measured. 
These protocols are motivated by the idea that a quantum annealer can provide more useful agorithmic tools if it is able to incorporate known information, either from classical algorithms, or from previous annealing runs \cite{chancellor17b}.
Since the introduction of the reverse annealing feature, there have been many experimental results based on this protocol showing improvements over traditional forward annealing.
These include: quantum simulation of the Kosterlitz-Thouless phase transition \cite{King2018a} which would not have been possible with traditional forward annealing; experiments showing that seeding the protocol with the result of a classical greedy search may improve portfolio optimisation \cite{Venturelli2019a}; evidence that repeated reverse anneals can improve performance in matrix factorization, \cite{Ottaviani2018a,Golden2020a}; and evidence that reverse annealing calls can improve genetic algorithms \cite{King2019a}. 

While these results are highly promising, they all focus on the improvements which reverse annealing can provide, rather than verifying that it actually searches locally, as predicted. 
In this paper we fill this gap by verifying that reverse annealing does indeed perform the kind of local search predicted in \cite{chancellor17b}.
The reverse annealing protocol, in the form in which it has been implemented on these devices, fundamentally relies on dissipation to find low energy states.
If all noise was removed and these protocols were run adiabatically starting from a unique classical basis state, they would return to the state in which they were initialized.
If run non-adiabatically, the protocol would incorporate nearby states, but there is no apparent mechanism for such a protocol to preferentially reach higher quality solution states, as there is for other non-adiabatic protocols \cite{Callison20a,crosson2020prospects}.
There are other biased search protocols which work in the adiabatic limit \cite{Perdomo-Ortiz11,Duan2013a,Grass19a}. 
Recent work \cite{Callison20a} suggests that the protocol explored in \cite{Duan2013a,Grass19a} will also perform well far from the adiabatic limit, in the rapid quench regime (fully coherent).

Since dissipation plays such an important role in the reverse annealing protocols implemented on D-Wave QPUs, they are a natural tool for studying the effect of noise.
The underlying protocol is very similar to the one used in macroscopic resonant tunnelling experiments \cite{Harris08a} to characterize noise on flux qubits. %
Our experiments reported here are not intended to examine single resonant tunnelling events, rather, they test the range of the local search which the annealer performs.  What makes our experiments particularly interesting is that we had access to two different QPUs which had different noise levels, allowing us to compare the performance between the two.
What we find is that, when the noise is reduced, the effective range of the local search is increased. 
Long range quantum tunnelling is a key feature of quantum annealers, that may one day allow them to obtain a quantum advantage. 
The increased range we observe is a sign of the beneficial effects of noise reduction in quantum annealers, even while thermal fluctuations play an important role in their operation.
It also provides an explanation for the improved performance on the lower noise QPU reported in \cite{noise_reduction_spin_glass_whitepaper}.

\section{Experimental setup}

\begin{figure}
\begin{centering}
\includegraphics[width=0.5\columnwidth]{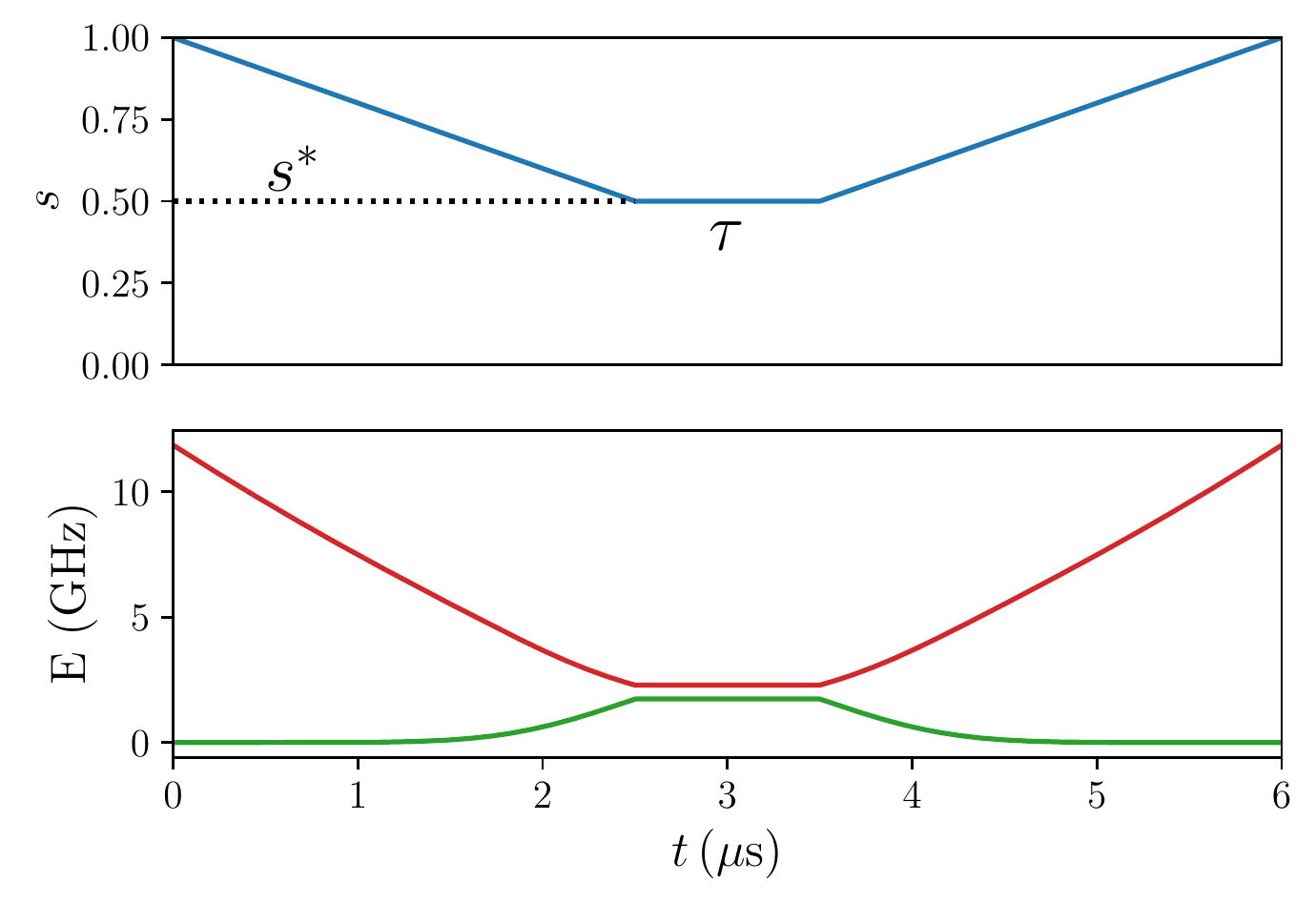}
\caption{\label{fig:reverse_anneal_protocol} 
Top: Plot of $s$ versus time for a reverse annealing protocol in which the device is annealed at the maximum allowed rate (which would take $5$ $\mu$s to traverse from $s=0$ to $s=1$), with a hold time $\tau$ of $1$ $\mu$s and $s^*=0.5$. Bottom: The $A$ and $B$ energy scales of this protocol performed on the low noise QPU model (see Fig.~\ref{fig:combined_schedule_plot} and associated discussion in the main text). The red (dark grey in print) curve starting from around $10$ GHz is $B$, the green (light grey in print) curve starting at around $0$ GHz is $A$. 
}
\end{centering}
\end{figure}

The D-Wave Systems programmable annealers that we used have a reverse annealing capability that works as follows. The annealing parameters $A$ and $B$ are defined to be functions which depend on a single parameter $s$, which is in turn time dependent.  The functional dependence of $A(s)$ and $B(s)$ is known \footnote{A table of parameter values at different values of $s$ for each device is provided to all QPU users.} but is non-linear and dependent on the physical device parameters. A key feature is that $A$ and $B$ are both monotonic in $s$, decreasing and increasing respectively, $A(s_1)>A(s_2)$ and $B(s_1)<B(s_2)$, if $s_1>s_2$. 
An initial state is pre-programmed into the annealer with control parameter $s=1$.
The control parameter $s$ is then linearly reduced until a value $s=s^*$ is reached, where the annealing protocol is paused for a period $\tau$. 
The protocol is finished by annealing back to $s=1$, and read out is performed normally. 
An example of a reverse annealing protocol is depicted in Fig.~\ref{fig:reverse_anneal_protocol}. 

These experiments were performed in a period of 2019 during which a new model of QPU was being introduced, which had been re-engineered to reduce the noise. 
During this period, the older, higher noise model was still available to use, allowing a unique opportunity to experimentally study the effect of noise on these devices. 
\begin{figure}
\begin{centering}
\includegraphics[width=0.5\columnwidth]{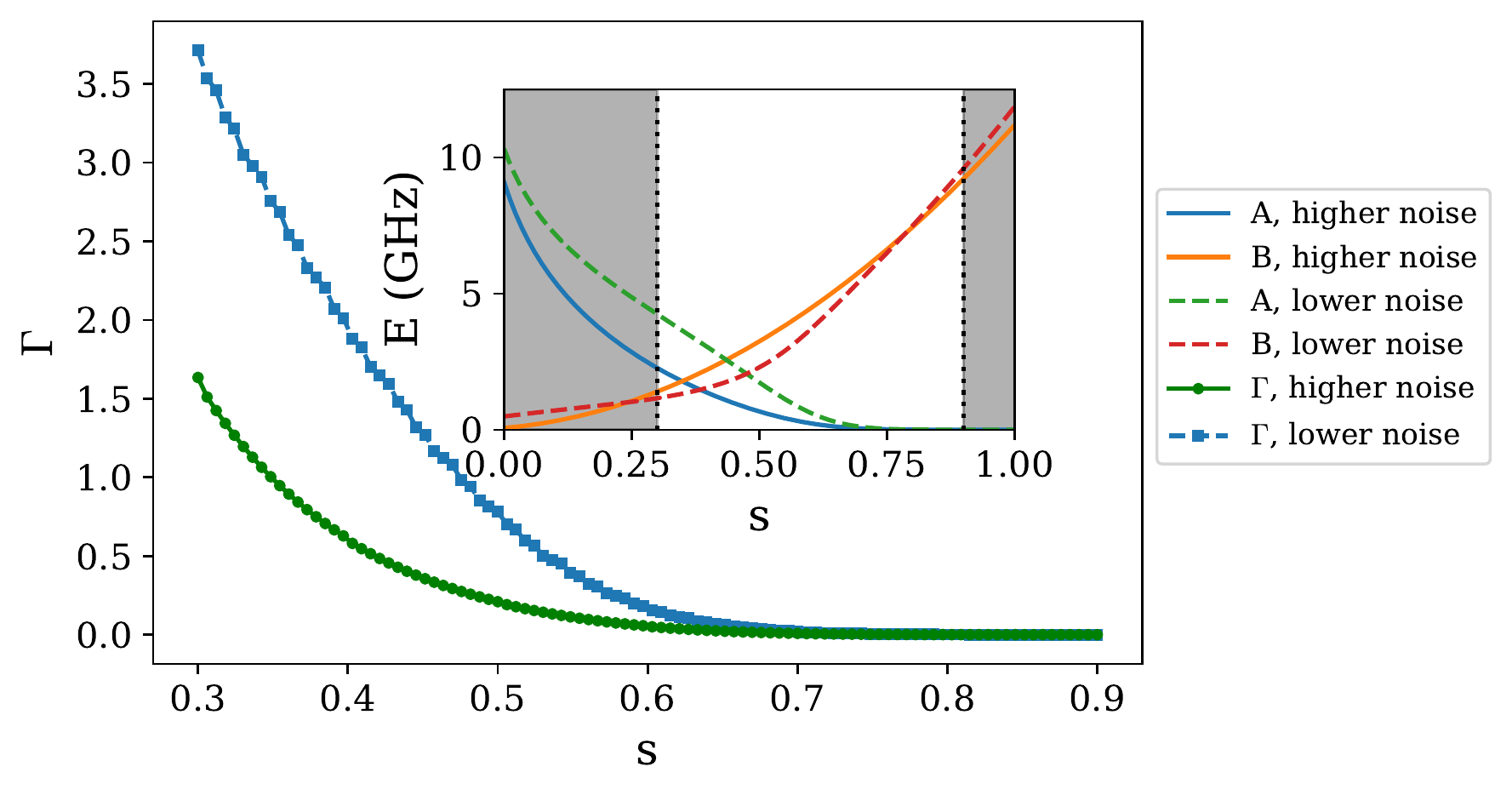}
\caption{\label{fig:combined_schedule_plot}
Ratio $\Gamma(s)=A(s)/B(s)$ for the high and low noise QPU at different values of $s$ within the range of interest. Blue squares (green circles) represent the different values of $\Gamma^*=\Gamma(s^*)$ used for the low (high) noise QPU. Inset: $A$ and $B$ curves for the higher (solid lines with $A$ in blue and $B$ in gold, both grey in print) and lower (dashed lines with $A$ in green and $B$ in red, both grey in print) noise QPUs ($A$ and $B$ curves can be distinguished in print either by matching the greyscale to the legend, or by recognizing that $A$ decreases while $B$ decreases). Greyed out areas fall outside of the range of interest, either because $\Gamma$ is very high and the system decays to its global ground state very rapidly, or because $\Gamma$ is very low and the dynamics are effectively frozen.
}
\end{centering}
\end{figure}
As depicted in the inset in Fig.~\ref{fig:combined_schedule_plot}, the high and low noise QPUs have different annealing schedules, so directly comparing their behaviour at the same value of $s$ does not provide a fair comparison. 
We instead define a single parameter $\Gamma(s)=A(s)/B(s)$ which encodes the ratio of fluctuation terms to problem Hamiltonian terms: this allows the results from the two devices to be compared on an equal footing. We correspondingly define $\Gamma^*=\Gamma(s^*)$.
We do make one exception to this rule in Fig.~\ref{fig:low_high_loc_barrier_sStar}, where we plot versus $s^*$ to make it clear what happens at high values of this parameter which are too squeezed to be readily visible on a plot versus $\Gamma^*$.

\subsection{Energy landscape design}

\begin{figure}
\begin{centering}
\includegraphics[width=0.5\columnwidth]{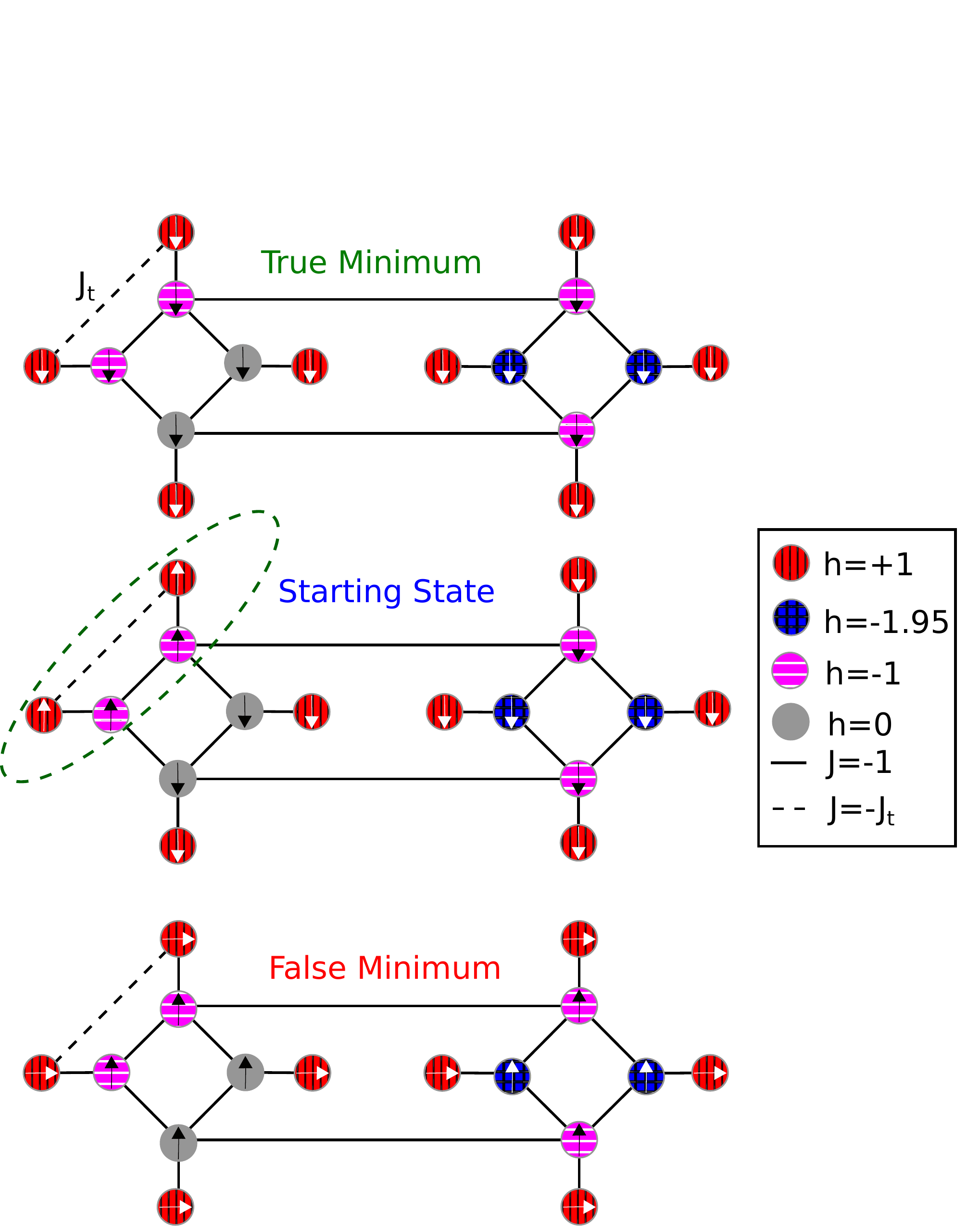}
\caption{\label{fig:three_states} 
The Hamiltonian used in the experiment and the three relevant sets of states, the true minimum (top), the starting state (middle), and the manifold of states which comprise the false minimum (bottom). Solid edges represent ferromagnetic couplings of unit strength ($J=-1$), the dashed edge represents a tunable ferromagnetic coupling with strength $|J|=J_t$. The red (grey in print) vertically hashed circles on the outer ring represent qubits with fields $h=+1$, the four vertically hashed purple (grey in print) circles in the inner ring have $h=-1$, the two dark blue (dark grey in print) square hashed circles have $h=-1.95$, and the two light grey circles on the inner ring have $h=0$. Arrows indicate the state of the qubit, with up (down) arrows representing $\ket{0}$ ($\ket{1}$), horizontal arrows representing superpositions of computational basis states which may be measured as either $\ket{0}$ or $\ket{1}$.  The dashed oval is a guide to the eye to show which qubits are flipped between the starting state and the true minimum energy state.
}
\end{centering}
\end{figure}

Since the goal of this experiment is to study local searches of solution spaces of optimisation problems, we construct a Hamiltonian where the energy landscape has the following features:
\begin{enumerate}
\item a broad (in the sense that it contains many nearby states with similar energies) false energy minimum which will be found with high probability using a traditional forward annealing protocol;
\item a narrow (relative to the false minimum) true minimum;
\item a local energy minimum (starting state) which sits relatively near (in Hamming distance) to the true minimum and far from the false minimum;
\item controllable barrier height between the starting state and the true minimum.
\end{enumerate}
Hamiltonians with the first two of these features are already known, and have been used experimentally in \cite{Boixo2013a,dickson13a}. 
We adapt the Hamiltonian used in \cite{dickson13a} to include a local minimum for the starting state.  
The Hamiltonian we use, along with the starting state, the unique state defining the true minimum, and the manifold of states which comprise the false minimum, are all represented pictorially in Fig.~\ref{fig:three_states}. 

Transitioning from the starting state (Fig.~\ref{fig:three_states} middle) to the true minimum energy state (Fig.~\ref{fig:three_states} top) can be achieved by flipping the state of only four qubits. 
Furthermore, if the two qubits on the outer ring are flipped first, the system only needs to pass through one state which has higher energy than the initial value. 
This higher energy state has the two qubits on the inner ring taking their initial value, but the two on the outer ring disagreeing, and this state will only have $2\,J_t$ more energy than the initial state.  For this reason, $J_t$ can be viewed as effectively acting as a barrier height between the initial state and the true minimum. 
If $J_t=0$, the true minimum can be reached without passing through any higher energy states. The size of the $J_t$ coupling will also have a minor effect on the ``width'' of the broader false minimum, by removing states where two of the spins disagree from the ground state manifold, and separating it based on whether the two spins it couples are up or down. However, this minimum is still much broader than the true minimum in all cases.

On the other hand, to reach the false minimum (Fig.~\ref{fig:three_states} bottom) from either the starting state or the true minimum, six spins on the inner rings need to be flipped, and the system must pass through states with a higher energy than the initial value regardless of the value of $J_t$. 
Compared to the transition between the starting state and the true minimum, more bit flips are required, and the system needs to pass through more than one state with a higher energy than the starting state. 
It is thus fair to characterize the transition between the starting state and the false minimum as being longer range (in Hamming distance) than the transition to the true minimum.

From summing the energy contributions from fields and couplers (and assuming that the two qubits coupled by the coupler with strength $J_t$ take the same value in the case of the false minimum), we see that the expectation value of the true and false minimum with respect to the dimensionless Hamiltonian $H_{\mathrm{prob}}$ 
only differs by $0.2$.
This implies that an avoided crossing between the true and false minima will occur relatively late in a traditional anneal, when the transverse field is very weak and the time scale for resonant tunnelling will be very long. 
This is by design, since our experiment is intended to study local search around the starting state, not resonant tunnelling between the two minima.
\begin{figure}
\begin{centering}
\includegraphics[width=0.5\columnwidth]{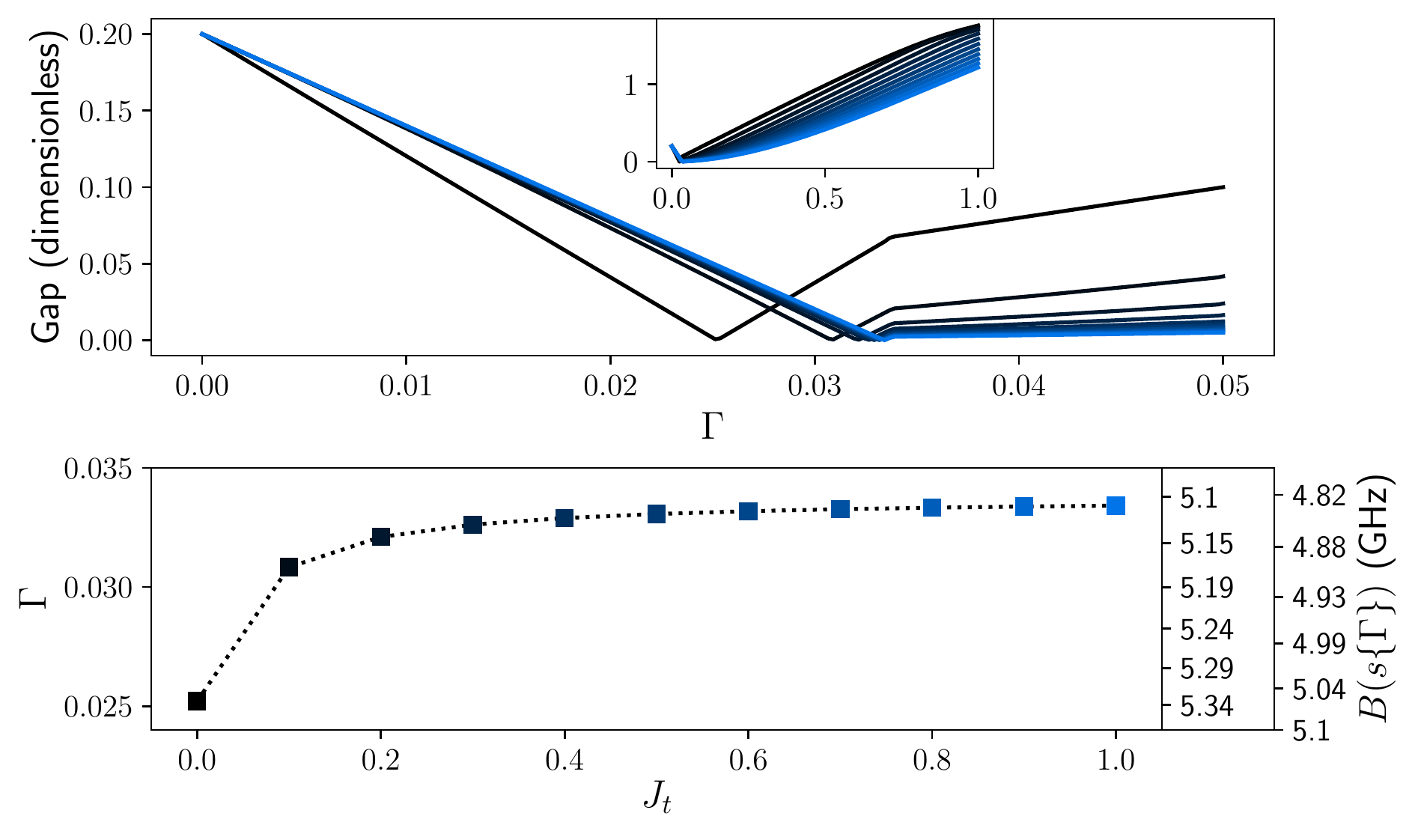}
\caption{\label{fig:Gadget_avoided_crossing} Upper: dimensionless gap versus $\Gamma$, for different values of $J_t$, as colour coded (greyscale coded in print) in the lower plot. Inset: same figure  plotted on a larger scale. Lower: position of the crossing versus $J_t$, the dimensionless ratio $\Gamma$ is on the left axis, while the values of $B(s\{\Gamma\})$ corresponding to these $\Gamma$ values for the different QPUs is given on the right axes, with the inner (outer) axis representing the lower (higher) noise model.}
\end{centering}
\end{figure}
We can confirm this numerically, to rule out the possibility of resonant tunnelling between the true and false minima. 
Fig.~\ref{fig:Gadget_avoided_crossing} depicts the energy gap between the ground and first excited state around the avoided crossing for different values of $J_t$. 
Due to the closeness of the avoided crossing, we were not able extract the exact gap values, but were able to numerically upper bound the minimum dimensionless gap to be of the order $10^{-9}$, using bisection based on numerical derivatives,
for all values of $J_t$. 
This implies that the gap will be of the order 
$10^{-9} B(s\{\Gamma\})$, where $s\{\Gamma\}$ is defined as the value of $s$ corresponding to $\Gamma$. 
The monotonicity of $A(s)$ and $B(s)$ implies that $\Gamma(s)$ will also be monotonic, and therefore $s\{\Gamma\}$ will be uniquely defined. 
From Fig.~\ref{fig:Gadget_avoided_crossing}, we see that $B(s\{\Gamma\})$ is of the order of GHz around the crossing point, for both the higher and lower noise QPUs. 
This means that the timescale related to resonant tunnelling can be lower bounded to a range of hundreds of milliseconds to seconds. 
Since the longest hold time used in our experiments ($100$ $\mu$s) is at least $1,000$ times smaller than this lower bound, we can safely conclude that resonant tunnelling directly between the true and false minimum is not playing a meaningful role in our experiments. The numerical methods used to calculate these values are discussed in detail in section \ref{sec:methods}.
This analysis does not exclude other relevant mechanisms, including: non-resonant tunnelling at larger $\Gamma$ values; the indirect thermally assisted processes observed by \citeauthor{dickson13a} \cite{dickson13a}; or other complex processes involving more than two energy levels.  
Ruling out direct resonant tunnelling is important, because it provides a contrast between our experiment and \cite{low_noise_whitepaper}, where reduced noise on the QPU leads to an increase in single and multi qubit macroscopic resonant tunnelling rates.

\section{Results}

\subsection{Local search by reverse annealing}

\begin{figure}
\begin{centering}
\includegraphics[width=0.5\columnwidth]{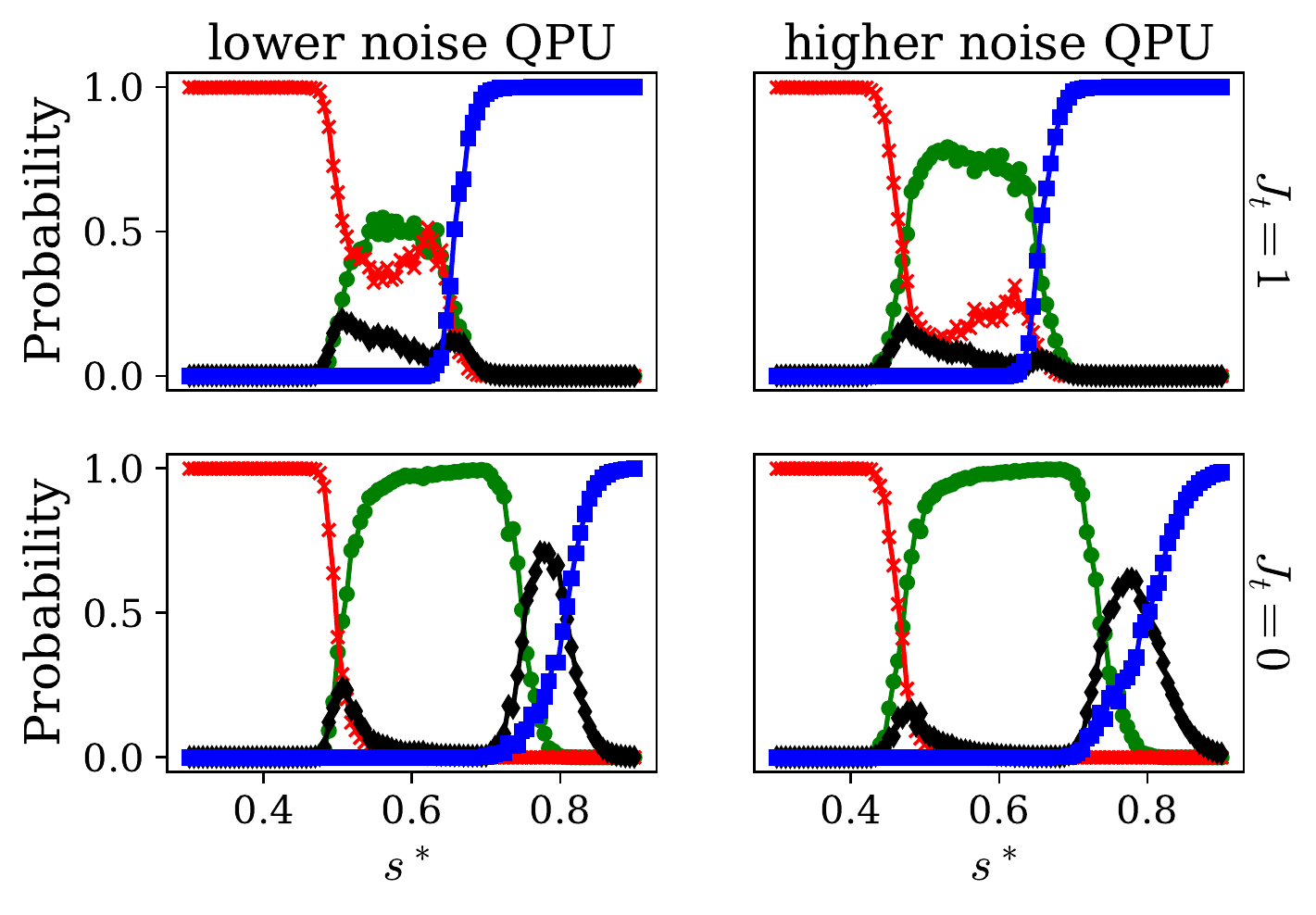}
\caption{\label{fig:low_high_loc_barrier_sStar}
Probability of the system to be in different states at the end of the anneal versus $s^*$ for lower and higher noise QPUs and different local barrier heights set by $J_t$. Top row: $J_t=1$, bottom row: $J_t=0$, left (right) column: lower (higher) noise QPU. Probability to be in the initial state (blue, grey in print, squares), the true minimum energy state (green, gray in print, circles), the false minimum energy state (red, grey in print, crosses), and any other state (black diamonds). All data in these plots were taken with a hold time of $\tau=5$ $\mu$s. 
Note that the schedules are different for the two QPUs, so a direct comparison at the same value of $s^*$ is not immediately meaningful; this plot uses $s^*$ rather than $\Gamma^*$ so that the behaviour at large $s^*$ is more readily visible. Statistical error bars are smaller than the symbols.
}
\end{centering}
\end{figure}

We first show that the results from the experiments reported here demonstrate that the reverse annealing protocol does perform searches locally in solutions space, as predicted by \citeauthor{chancellor17b} \cite{chancellor17b}. 
It is valuable to directly verify this behaviour experimentally in small systems, to provide a firm foundation for building more complex algorithms. 
In this setting, what we expect is that for intermediate values of $s^*$ (equivalently, intermediate $\Gamma^*$), the system should find the true minimum with reasonable probability. 
In contrast, for too large a value of $s^*$ (small $\Gamma^*$), the system will remain frozen in its initial state, and for too small a value of $s^*$ (large $\Gamma^*$), the system will search the solution space globally and become trapped in the false minimum. This is to be expected because of how the problem is designed; for too small a value of $s^\star$ the second half of the reverse annealing protocol becomes indistinguishable from forward annealing.
Fig.~\ref{fig:low_high_loc_barrier_sStar} shows that this is indeed the case for both QPUs.
For intermediate values of $0.5\lesssim s^*\lesssim 0.7$, the true minimum (green) is found preferentially.
Above and below this range, the initial state (blue) and false minimum (red) are observed respectively, with near unit probability.
If $J_t=0$ then no energy barrier needs to be overcome to transfer from the starting state to the true minimum and therefore nearly all of the probability amplitude is able to be transferred. 
A large peak in the probability to be in a state which is neither the true minimum, the false minimum, nor the starting state can also be observed for some values of $s^*$ at the edges of the range, $s^*\simeq 0.5$ and $s^*\simeq 0.7$. 
This is likely due to slow dynamics leading to the system becoming stuck in a state intermediate between the starting state and the true minimum.

A larger barrier $J_t=1$ seems to trap the amplitude in the initial state until a relatively smaller value of $s^*$ is used. This leads to less than $100\%$ of the probability in the true minimum since at smaller $s$ (larger $\Gamma$), the amplitude is able to tunnel to the false minimum more effectively.  This feature is more pronounced in the lower noise QPU.

\begin{figure}
\begin{centering}
\includegraphics[width=0.5\columnwidth]{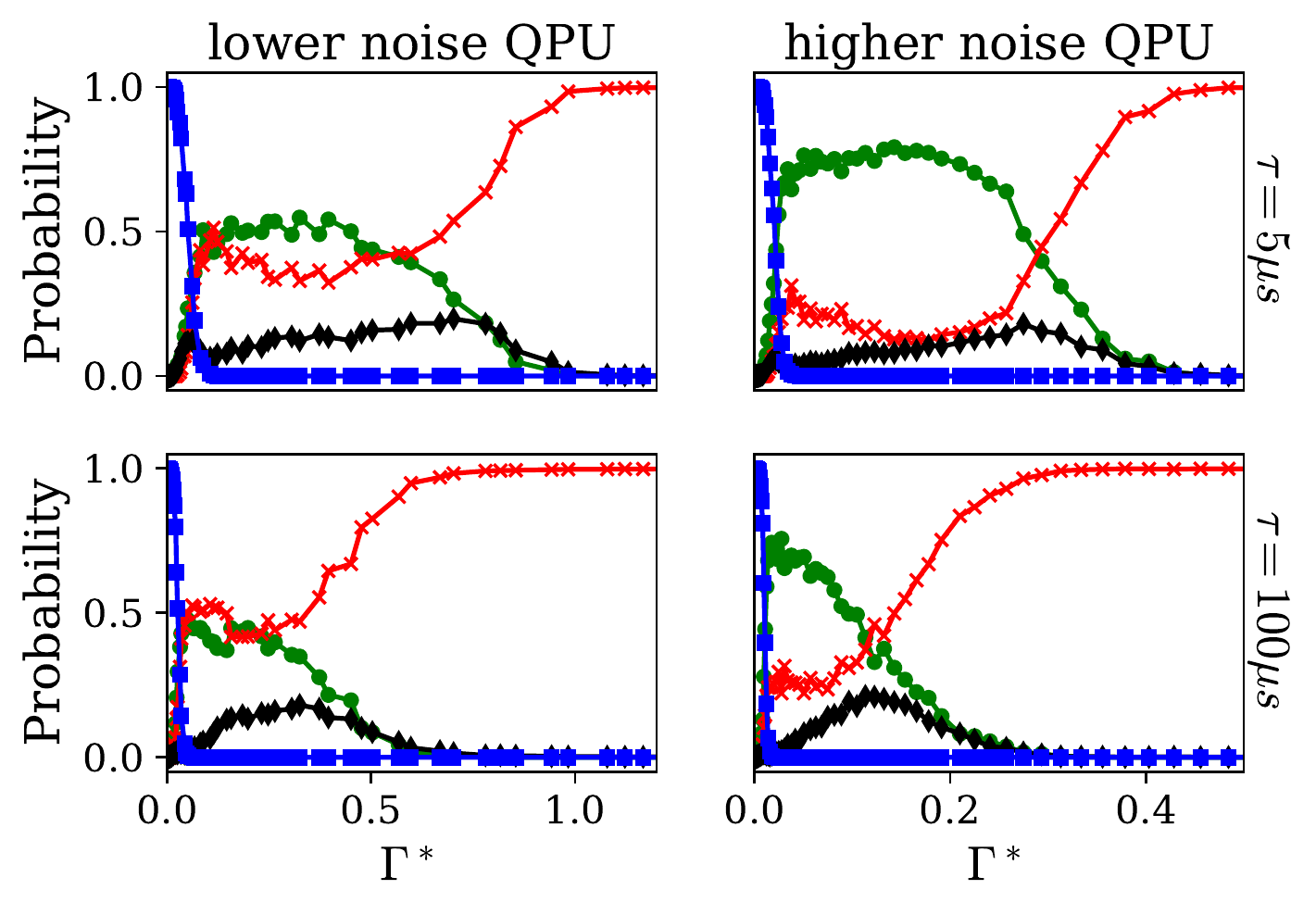}
\caption{\label{fig:low_high_tau_Gamma}
Probability of the system to be in different states versus $\Gamma^*$ for lower and higher noise QPUs and two different hold times $\tau$. Top row: $\tau=5$ $\mu$s, bottom row: $\tau=100$ $\mu$s, left (right) column: lower (higher) noise QPU. Probability to be in the initial state (blue, grey in print, squares), the true minimum energy state (green, gery in print, circles), the false minimum energy state (red, grey in print, crosses), and any other state (black diamonds).  All plots here are for $J_t=1$. Statistical error bars are smaller than the depicted symbols.
}
\end{centering}
\end{figure}

The plots in Fig.~\ref{fig:low_high_loc_barrier_sStar} are all for a fixed hold time $\tau=5$ $\mu$s (see Fig.~\ref{fig:reverse_anneal_protocol}).
The effect of varying the hold time $\tau$ is shown in  Fig.~\ref{fig:low_high_tau_Gamma}, which is plotted against $\Gamma^*$ instead of $s^*$. 
The top row shows the same data as the top row of Fig.~\ref{fig:low_high_loc_barrier_sStar}, while the bottom row is for hold time $\tau=100$ $\mu$s.
Longer hold times reduce the value of $\Gamma^*$ (equivalently, increase $s^*$) above which the system is no longer found in the global minimum, and also reduce the value of $\Gamma^*$ (increase $s^*$) 
at which the system leaves the initial state.
Both of these effects make intuitive sense, as the system is more likely to transition to a lower energy state given more time. 
Two further observations are also worth making. 
Firstly, the QPU with higher noise levels makes these transitions at lower values of $\Gamma^*$ than the QPU with lower noise levels (note the different scales for $\Gamma^*$ in Fig.~\ref{fig:low_high_tau_Gamma}).
This again makes intuitive sense. 
The computation is driven by thermal dissipation mediated by coupling between the substrate of the QPU and the qubits. 
This coupling will be stronger in the higher noise QPU, and therefore the timescales associated with dissipation will be shorter, allowing dissipation to happen at lower values of $\Gamma^*$ for the same hold time $\tau$. 
Secondly, the peak value of the probability to be found in the true minimum is very similar for both hold times.
This suggests that, at relatively low $\Gamma^*$ (but still large enough to leave the initial state), the timescale associated with transferring from the true minimum to the false minimum is much longer than even the longest hold time of $\tau=100$ $\mu$s. 
Thus, we have a picture of fast transfer out of the initial state, followed by relatively slow transfer between the other minima. 
Furthermore, the initial transfer strongly favours the nearby true minimum over the false minimum further away. 
This demonstrates that the reverse annealing protocol does indeed search the solution space locally, validating the conceptual argument that the performance increases seen in reverse annealing protocols are due to an ability to preferentially search locally in promising areas of the solution space.

\subsection{Search range and noise levels}

Having demonstrated that reverse annealing performs a local search on QPUs with both higher and lower noise levels, we now look at the differences in the behaviour of the two devices. 
We have already observed two basic effects. 
Firstly, the effective timescale for transferring between states is shorter for the higher noise QPU. 
Secondly, and more interestingly, we observe that 
the low noise system appears to have a relatively higher probability to transfer amplitude to the false minimum, which is further in Hamming distance from the initial state than the true minimum. 
Although in this engineered problem, this means that the probability of ``success'' is lower in our experiment, it hints at something very important, namely that the transfer between states has a longer range character in the lower noise QPU. 
The use of quantum coherence for long range exploration of a solution space is one of the attributes which should be able to give quantum annealing an advantage over classical methods.
It is thus worth examining the search range on both QPUs more carefully.

\begin{figure}
\begin{centering}
\includegraphics[width=0.5\columnwidth]{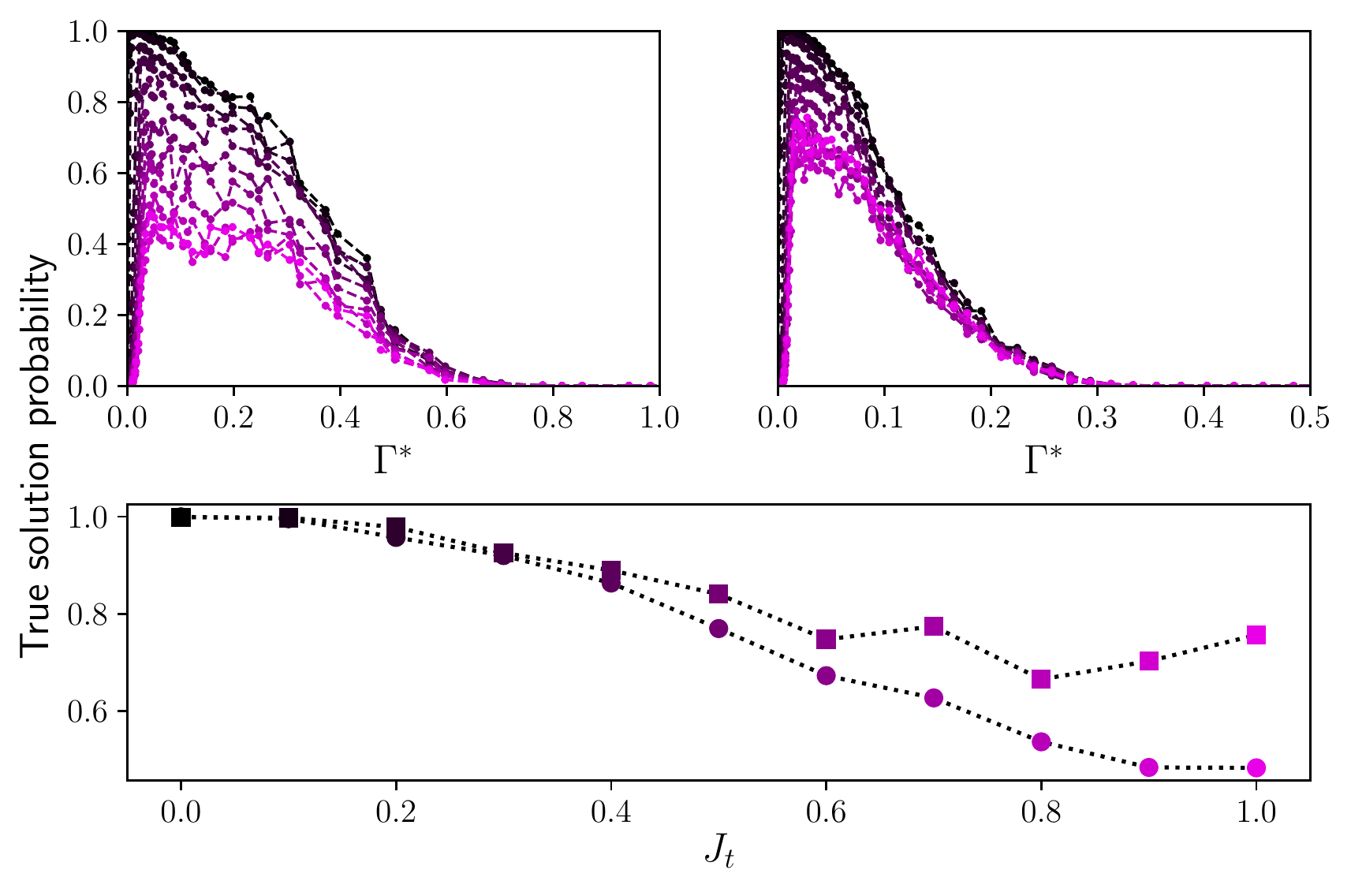}
\caption{\label{fig:Jt_peak_trueMin}
Top: probability of the system being found in the true minimum for different values of $\Gamma^*$, left (right) is lower (higher) noise QPU.  Both plots are for a hold time of $\tau=100$ $\mu$s. Values of $J_t$ are colour coded (greyscale coded in print) to correspond to the bottom plot. Bottom: peak probability of being in the true minimum for high noise (squares), and low noise (circles) QPUs. Statistical error bars on the upper plots are smaller than the symbols.
}
\end{centering}
\end{figure}

\begin{figure}
\begin{centering}
\includegraphics[width=0.5\columnwidth]{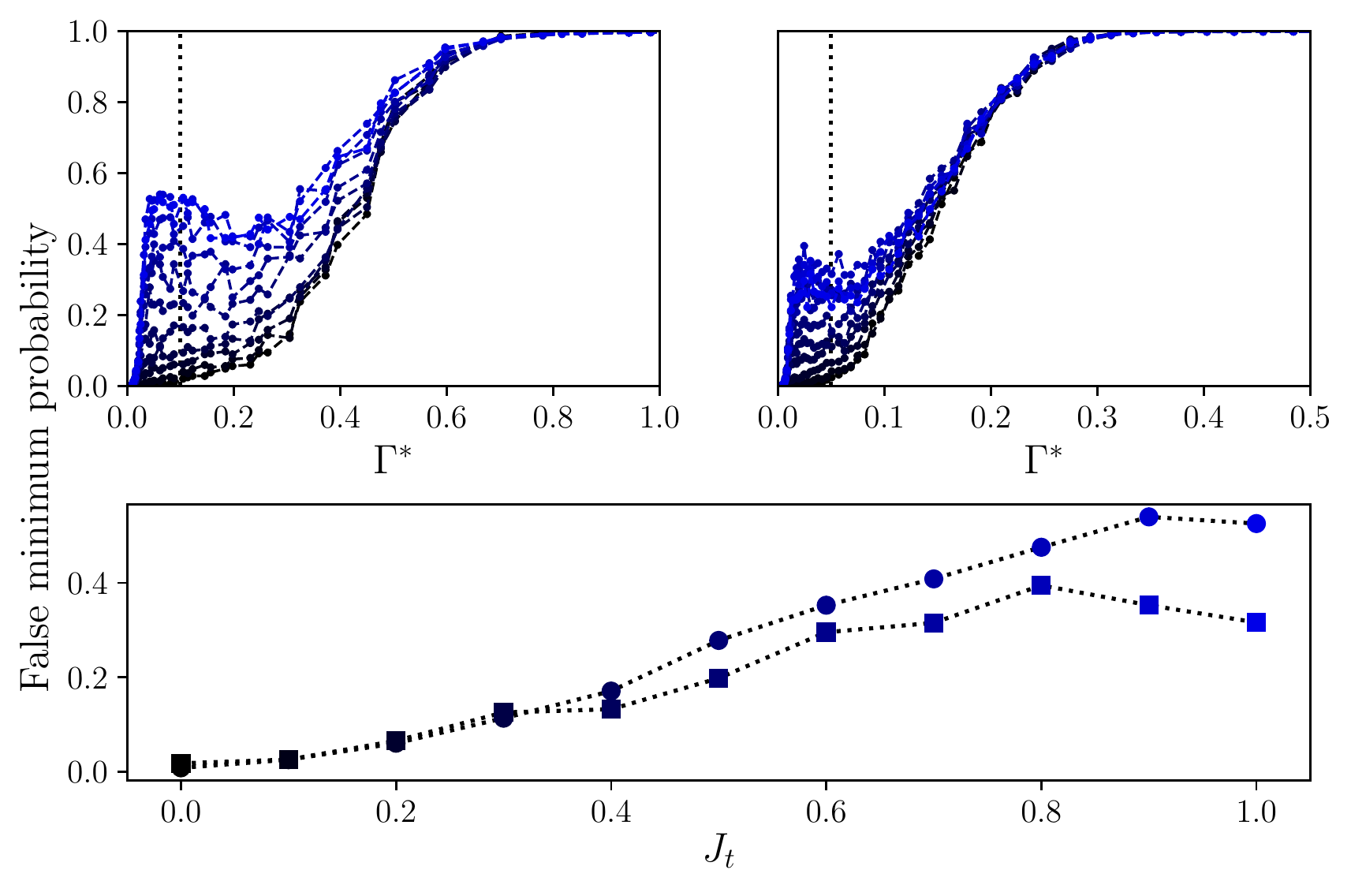}
\caption{\label{fig:Jt_falseMin.pdf}
Top: probability of the system being found in the false minimum for different values of $\Gamma^*$, left (right) is lower (higher) noise QPU.  Both plots are for a hold time of $\tau=100$ $\mu$s. Values of $J_t$ are colour coded (greyscale coded in print) in a way which corresponds to the bottom plot. Bottom: maximum probability to be found in the false minimum for $\Gamma\le 0.1$ ($\Gamma\le 0.05$) on the lower (higher) noise QPU. As the dashed lines on the top plots show, these values were chosen to capture the initial plateau, before complete transfer into the false minimum.
}
\end{centering}
\end{figure}

First, we examine what happens as we change $J_t$, the strength of the barrier between the initial state and the true minimum. 
As Fig.~\ref{fig:Jt_peak_trueMin} illustrates, the probability to reach the true solution (and therefore not end up in the false minimum) is roughly the same for both QPUs if the barrier between the initial state and true solution state is small. 
The behaviour of the two QPUs starts to diverge around $J_t=0.4$, with the higher noise QPU showing a lower probability of finding the true minimum for  $J_t>0.4$.
This intuitively makes sense: without a barrier preventing rapid decay into the true minimum, this decay will happen before $\Gamma$ is large enough to reach the false minimum. 
As Fig.~\ref{fig:Jt_falseMin.pdf} demonstrates, the state is tunnelling into the false minimum preferentially on the lower noise device, indicating a longer range character to the search.

\begin{figure}
\begin{centering}
\includegraphics[width=0.5\columnwidth]{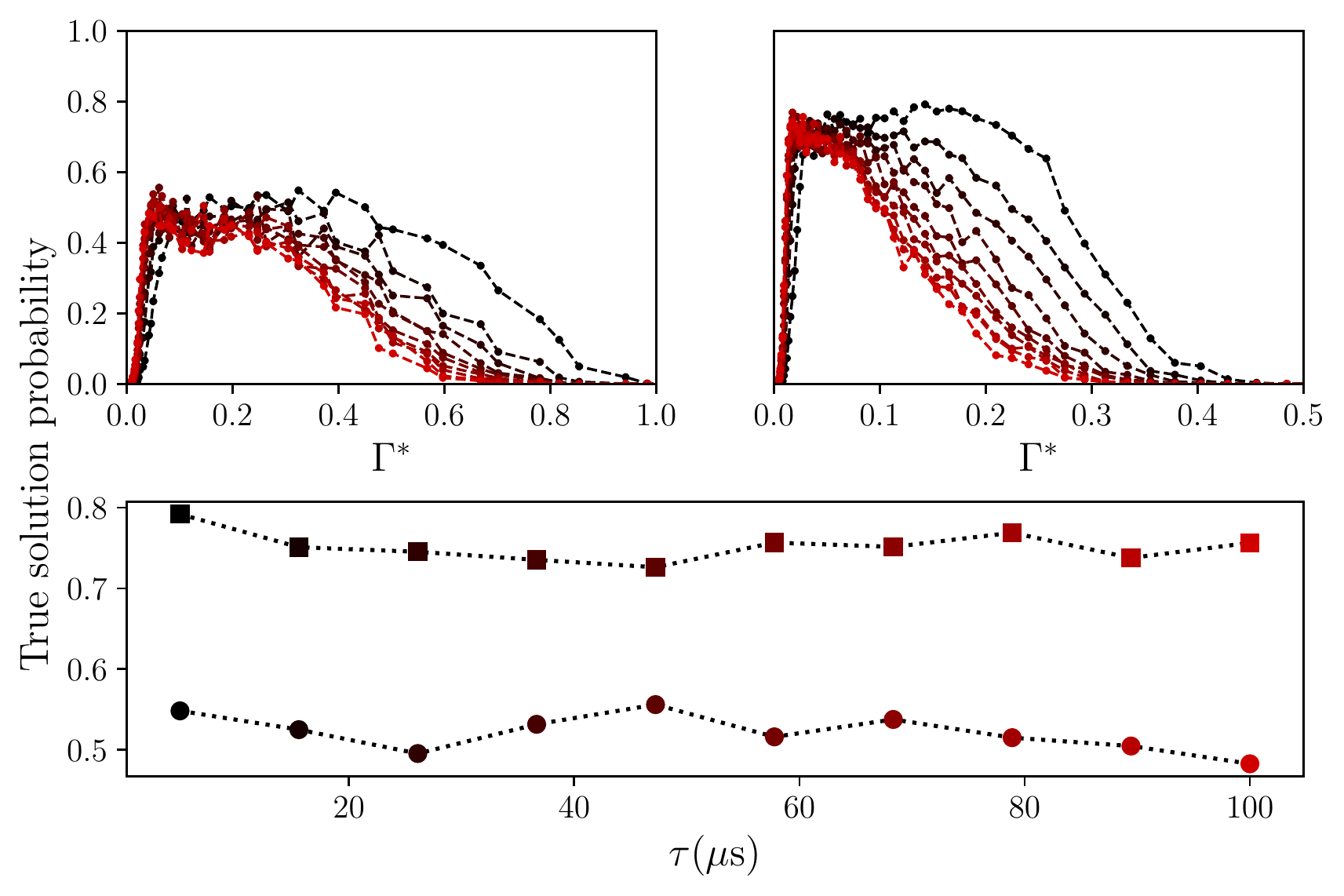}
\caption{\label{fig:tau_peak_trueMin}
Top: probability of the system being found in the true minimum for different values of $\Gamma^*$, left (right) lower (higher) noise QPU. Both plots are for a barrier strength of $J_t=1$. Values of $\tau$ are colour coded (greyscale coded in print) in a way which corresponds to the bottom plot. Bottom: peak probability of being in the true minimum for high noise (squares), and low noise (circles) QPUs. Statistical error bars on the upper plots are smaller than the symbols.
}
\end{centering}
\end{figure}

We now more carefully examine the effect of the hold time $\tau$ on the peak probability to be found in the true minimum. 
As the data in the previous section suggest, the peak value seems to be unaffected by the hold time, consistent with a separation of time scales between tunnelling out of the initial state and tunnelling between the true and false minima. 
Fig.~\ref{fig:tau_peak_trueMin} confirms that at least for a range of $\Gamma$ values right around the initial peak in true minimum probability, this separation of timescales is a valid assumption, with very little effect on the peak value as hold time is increased.

\begin{figure}
\begin{centering}
\includegraphics[width=0.5\columnwidth]{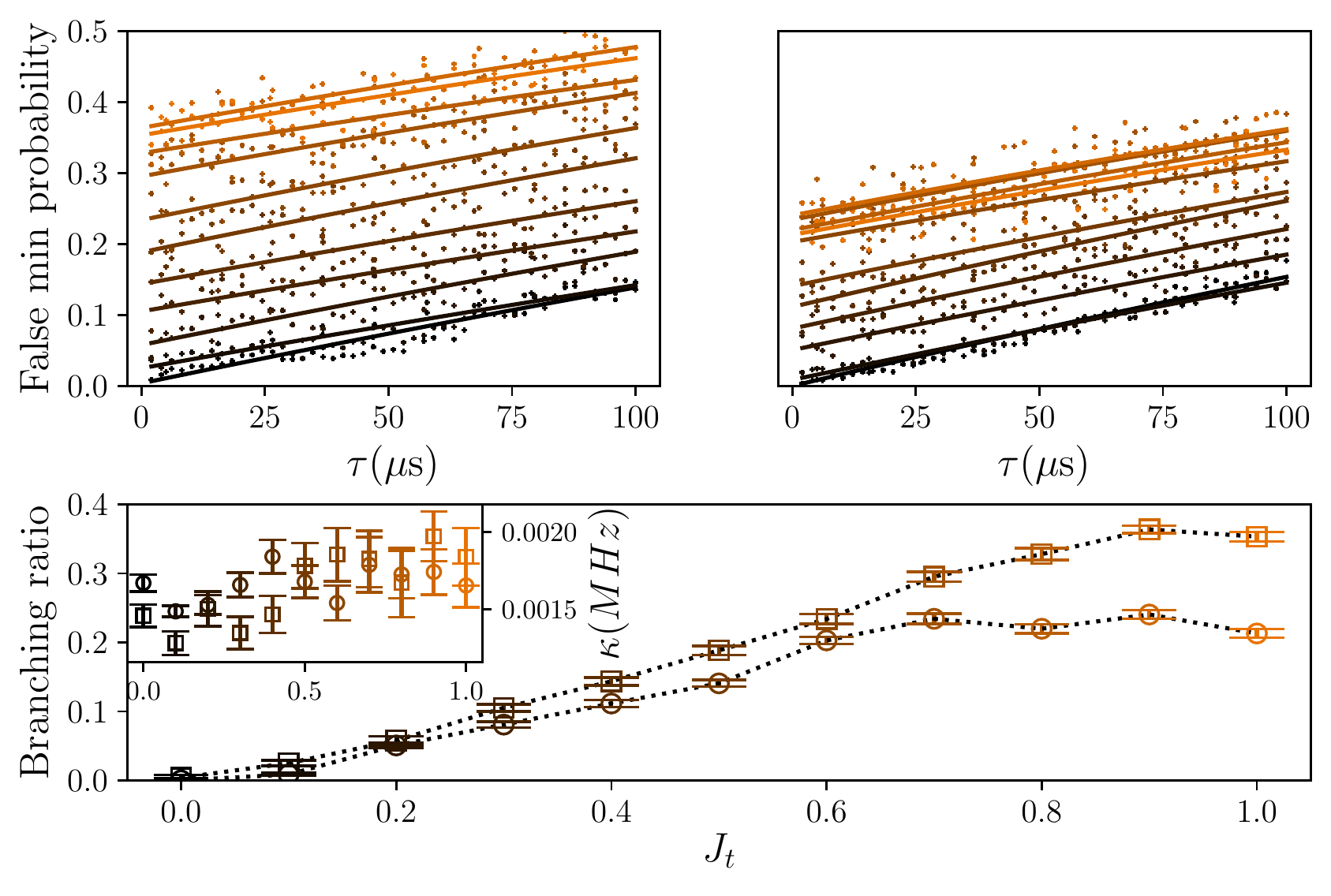}
\caption{\label{fig:branching_ratio_falseMin}
Top: Fits to extract the branching ratio from the initial state to the false minimum  at $s^*=0.57$ corresponding to $\Gamma^*=0.31$ for the lower noise QPU and $\Gamma^*=0.089$ for the higher noise QPU. Left (right) is lower (higher) noise QPU. Lines represent the fits, while points represent the individual measurements values of $J_t$ are colour coded (greyscale coded in print) in a way which corresponds to the bottom plot. Bottom main: Extracted values of the branching ratios for the higher and lower noise QPUs ($R_{\mathrm{false}}$ from Eq.~\ref{Eq:branching_fit}). Bottom inset: Extracted values of $\kappa$. Circles (squares) represent higher (lower) noise QPU. Error bars are the square root of the variance (i.e. the diagonal elements of the covariance matrix) for each parameter in the non-liner fit. Note that while the statistical error bars on the upper plots are smaller than the symbols, there is clearly some more systematic error present, possibly related to long timescale noise on the device.
}
\end{centering}
\end{figure}

Based on this separation of timescales we can create a simplified model of the dynamics, in which transfer out of the starting state to the two minima occurs on a very fast timescale, but then transfers from the true minimum to the false minimum occurs on a slower timescale. In this model, we consider an initial transfer which occurs instantaneously and leaves the system in the false minimum with probability $R_{\mathrm{false}}$, and the true minimum with probability  $\simeq(1-R_{\mathrm{false}})$. After this initial transfer, there is a slow decay from the true minimum into the false minimum, this decay is responsible for the difference in probabilities at different values of hold time $\tau$.
Given this separation of timescales, we can directly estimate the branching ratio $R_{\mathrm{false}}$ from the initial state into the false minimum for both the lower and higher noise QPUs. 
To do this, we take data at many values of $\tau$ for $s^*=0.57$ on both the higher and lower noise QPUs.
The value of $s^*=0.57$ was chosen because on one hand even for the smallest $\tau$, this is well above the value of $\Gamma$ where there is any chance to remain in the initial state, so we are justified in approximating the initial transfer as instantaneous. On the other hand there is still significant probability to be in the true minimum even for the largest value of $\tau$, so the process of transfer between the true and false minima is well captured and can be accurately fit.
From these data, we are able to fit the following simplified model 
\begin{equation}
P_{\mathrm{false}}(\tau)=1-\left[1-R_{\mathrm{false}} \right] \exp(-\kappa \tau) \label{Eq:branching_fit}
\end{equation}
where $P_{\mathrm{false}}(\tau)$ is the probability of being in the false minimum, observed at a given hold time $\tau$. 
In this simplified model, there is one timescale $\kappa$ which determines the rate at which amplitude decays from other states (mostly the true minimum) into the the false minimum. 
As Fig.~\ref{fig:branching_ratio_falseMin} demonstrates, the branching ratio extracted by fitting does indeed show that the lower noise QPU favours longer range transport of probability amplitude, because it has a larger value of the branching ratio $R_{\mathrm{false}}$ for $J_t>0.4$, i.e., when the barrier height is significant.
Although the branching ratio depends on the barrier height, the decay rate $\kappa$ does not (within errors).
This is expected, since there is no obvious mechanism through which the barrier height should have  strong effect on the rate of transfer between the true and false minima.

\section{Methods \label{sec:methods}}

The experimental data presented in this paper were obtained using using the Matlab API to remotely access the QPUs on September 21$^{\mathrm{st}}$ and September 22$^{\mathrm{nd}}$ 2019. Numerical analysis and creation of plots was performed in Python \cite{van2003python} using the numpy \cite{oliphant2006guide}, scipy \cite{scipy}, and matplotlib \cite{hunter2007matplotlib} packages as well as Jupyter notebooks \cite{jupyter}. Data are publicly available \cite{search_range_data}. 

For each raw data point, $1,000$ annealing runs were performed. 
Multiple copies of the gadget were embedded into the QPU for each run, allowing all $11$ values of $J_t$ to be annealed simultaneously. 
A $16 \times 16$ chimera graph QPU could accommodate $128$ copies of the experimental Hamiltonian, if all qubits were functioning.
However, both devices contain a small number of qubits which are non-functional due to failed calibration.  For this reason, only $118$ total copies of the Hamiltonian were used per run for the higher noise QPU, and $116$ for the lower noise QPU. 
After dividing as evenly as possible over the $11$ values of $J_t$, this still leaves at least $10$ copies for each value of $J_t$ on each QPU, and therefore at least $10,000$ individual samples for each data point. 
The standard error for each data point is therefore \textbf{at most} $0.5 \% $ (this assuming the worst case scenario of a result which occurs $50\%$ of the time). 
As this is smaller than the symbols depicted in our plots, statistical error bars resulting from sampling errors have been omitted from all figures. 
Error bars calculated using different methods are depicted in some figures, as explained in the captions. 
The definitions of logical $1$ and $0$ were chosen randomly to avoid systematic errors from uniform fields (this process is sometimes referred to as gauge averaging).

Avoided crossing location and estimated gaps were calculated using numerical bisection on numerically calculated derivatives of the gap. 
To obtain the gap using reasonable computational resources, the \textbf{lobpcg} function (which can be found in \textbf{scipy.sparse.linalg}) was used to calculate the gaps. This function takes as inputs estimates of the eigenstates. We used the first $5$ eigenstates found at $\Gamma=1$, using the known state which comprises the true minimum as a sixth state. 
We have verified by plotting points near the gap location which our bisection has found, that it was not able to fully converge, but is still close enough to the avoided crossing that there is no visible difference between the location it found and the true avoided crossing on the scale plotted in Fig.~\ref{fig:Gadget_avoided_crossing}. 
However, this does mean that the gaps found by bisection should be treated as upper bounds, since they do not lie directly at the crossing. Since tighter bounds on these gaps would not change our interpretation of the experiments, we have elected not to pursue a higher quality estimate.

\section{Discussion and conclusions}

In this work, we have presented the results of a small system reverse annealing experiment using an engineered energy landscape, run on D-Wave QPUs which have different noise levels. 
Our experiment demonstrates the operating principles behind the reverse annealing protocol and verifies that it does indeed perform a local search as predicted in \cite{chancellor17b}. 
To our knowledge, this is the most direct demonstration of this behaviour to appear in the literature. 
More importantly, due to the availability of QPUs with two different noise levels, these experiments have allowed us to examine the effect of reducing the noise on these devices. 
Crucially, in addition to the expected reduction in dissipation driven state transfer rates, we find that reducing the noise levels results in local searches which are effectively longer range. 
Since long range quantum tunnelling is necessary for annealers to obtain a quantum advantage, the increase in range which we observe is a positive sign for improving the performance of these devices by reducing noise levels. 
In the interest of transparency and because re-analysis of our data may be important in understanding open quantum system models of flux qubit quantum devices, we have made our data publicly available \cite{search_range_data}. 

Reduced noise in QPUs leads to improved performance, as observed in \cite{noise_reduction_spin_glass_whitepaper}.
Our results suggest a potential mechanism for this performance improvement, 
namely, that noise reduction increases the range in solution space over which these devices are able to search. 
While it may seem obvious that reducing noise improves performance, it is actually far from trivial, as dissipation is the driving mechanism behind the currently implemented reverse annealing protocols, and thermal noise has been observed to improve performance in general in QPUs \cite{dickson13a}. 
It is therefore not clear whether the best performance could be obtained by removing all coupling to the thermal bath, or whether there is a ``sweet spot'' where a low (we would speculate still lower than the noise level of the current lowest noise QPUs) but still significant level of dissipation actually improves device performance. 

We explore this tradeoff experimentally, it would likely to be very fruitful to gain a fuller understanding both numerically and theoretically. While such analysis is beyond the scope of the current paper, we have provided an important first step toward a fuller understanding of how dissipative quantum annealers explore the space of solutions. Since our data have all been made publicly available (see: \cite{search_range_data}), they can provide an important check for theoretical and numerical efforts to further this exploration. In particular, one interesting question raised by this work is whether the effect on search range seen here is captured by a noise model which assumes that dissipation and dephasing occur within the energy eigenbasis of the problem Hamiltonian, (as is often assumed when formulating a master equation) or whether a model which includes disruption to these eigenstates is needed. This understanding is not only important for developing appropriate numerical models, but also is important for gaining a theoretical perspective on how the noise can disrupt the computational tunnelling, whether the decreased search range is simply because faster energy loss leads to decay into nearer states, or whether the system is being disrupted in a more fundamental way in which tunnelling mechanisms are being fundamentally changed. Fully understanding this point is likely to require extensive numerical studies, and is therefore outside of the scope of the present work.

What we can definitely say from our experiments is that, for the system and problem size we have studied, the most recent noise reduction has improved an important aspect of the device performance, the range of the local searching. This suggests that further noise reduction is an interesting route towards device improvement.  Our results also motivate future work using larger problem sizes and different Hamiltonians, to  determine how general our results are, and whether there is a limit to the improvements noise reduction can deliver on QPUs of this design. Furthermore, it would be interesting to study the effect of varying temperature on the process we study here, which we were not able to do since temperature controls are not publicly accessible.
The branching ratio experiments reported here provide a set of tools to explore search range in dissipative quantum annealing systems. This is important because it allows a quantity which is likely to be of high importance for computation to be directly probed experimentally. These experiments occupy a middle ground between the low-level experiments such as macroscopic resonant tunnelling which directly characterize device properties and full computational benchmarks which directly evaluate performance. By illuminating computationally relevant aspects of the device behaviour without relying on the specific details of a problem (including the difficulties associated with developing problems which are computationally hard \cite{Katzgraber2014a}), the techniques developed here provide a powerful and relatively straightforward tool to understand and characterize them. Moreover, since our experiments can be performed using access which is typically given to the end users of quantum annealers, they provide an important tool to independently check and understand these devices, without being granted special low-level access by the manufacturer.

\section*{Acknowledgments}

 NC and VK were supported by EP/L022303/1 and impact acceleration funding associated with this grant. NC was supported by EPSRC fellowship EP/S00114X/1.

\bibliography{bibLibrary}  % bibtex entries in separate file for convenience

\end{document}